%% file: Paper.tex
\documentclass[conference]{IEEEtran}
\input{Preamble}

\IEEEoverridecommandlockouts\IEEEpubid{\makebox[\columnwidth]{ 978-1-6654-3540-6/22~\copyright~2022 IEEE \hfill} \hspace{\columnsep}\makebox[\columnwidth]{ }}
\newcolumntype{M}[1]{>{\centering\arraybackslash}m{#1}}
\begin{document}

\title{Demonstrating the Merits of Integrating Multipath Signals into 5G LoS-Based Positioning Systems for Navigation in Challenging Environments}

\author{ 
Qamar Bader\textsuperscript{\orcidicon{0000-0002-4667-1710}}$^{* \dag}$, 
Sharief Saleh\textsuperscript{\orcidicon{0000-0003-1365-417X}}$^{* \dag}$,
Mohamed Elhabiby\textsuperscript{\orcidicon{0000-0002-1909-7506}}$^{\ddag}$,
and Aboelmagd Noureldin\textsuperscript{\orcidicon{0000-0001-6614-7783}}$^{*\dag}$\\

\IEEEauthorblockA{
$^*$ Department of Electrical and Computer Engineering, Queen's University, Kingston, ON, Canada\\
$^\dag$ Navigation and Instrumentation (NavINST) Research Lab,\\ Department of Electrical and Computer Engineering, Royal Military College of Canada, Kingston, ON, Canada\\
$^{\ddag}$Micro Engineering Tech. Inc.\\
Email: \{qamar.bader, sharief.saleh, nourelda\}@queensu.ca}

\thanks{\hspace{-7pt}\rule{\linewidth}{0.4pt}}
\thanks{This work was supported by grants from the Natural Sciences and Engineering Research Council of Canada (NSERC) under grant number: ALLRP-560898-20 and RGPIN-2020-03900.}}

\maketitle

\begin{abstract}	
Constrained environments, such as indoor and urban settings, present a significant challenge for accurate moving object positioning due to the diminished line-of-sight (LoS)communication with the wireless anchor used for positioning. The 5th generation new radio (5G NR) millimeter wave (mmWave) spectrum promises high multipath resolvability in the time and angle domain, enabling the utilization of multipath signals for such problems rather than mitigating their effects. This paper investigates the benefits of integrating multipath signals into 5G LoS-based positioning systems with onboard motion sensors (OBMS). We provide a comprehensive analysis of the positioning system's performance in various conditions of erroneous 5G measurements and outage scenarios, which offers insights into the system's behavior in challenging environments. To validate our approach, we conducted a road test in downtown Toronto, utilizing actual OBMS measurements gathered from sensors installed in the test vehicle. The results indicate that utilization of multipath signals for wireless positioning operating in multipath-rich environments (e.g. urban and indoor) can bridge 5G LoS signal outages, thus enhancing the reliability and accuracy of the positioning solution. The redundant measurements obtained from the multipath signals can enhance the system's robustness, particularly when low-cost 5G receivers with a limited resolution for angle or range measurements are present. This holds true even when only considering the utilization of single-bounce reflections (SBRs).
\end{abstract}


\section{Introduction}
Accurate and reliable positioning of moving objects in indoor environments is a critical challenge for various applications, such as asset tracking, indoor navigation, and emergency response. Over the past decade, several indoor positioning systems have been proposed and implemented, using various technologies such as Wi-Fi \cite{wifi}, Bluetooth \cite{BLE}, and Ultra-Wideband (UWB) \cite{uwb}. However, each technology has its limitations in terms of accuracy, reliability, scalability, and cost \cite{indoorChallenges}. Recently, the emergence of 5G cellular networks has provided a new opportunity for indoor positioning, thanks to the support of advanced features such as directional antennas, beamforming, and multi-antenna arrays \cite{5GPositioning}. In addition, 5G networks have higher multipath resolvability, enabling the utilization of multipath signals that were previously difficult to extract in such environments. This makes 5G a promising candidate for indoor positioning systems, as it can provide more accurate and robust positioning solutions. The exploitation of multipath signals in indoor positioning systems leads to a decrease in 5G signal outages, thus enhancing the reliability and accuracy of the positioning solution. Furthermore, the additional measurements provided by the multipath signals can improve the robustness of the system, particularly in the presence of low-cost 5G receivers with a limited resolution for angle or range measurements.

The objective of this work is to examine the advantages of incorporating multipath signals into 5G-based integrated positioning systems for indoor environments. The integrated positioning system being analyzed in this paper is based on our previous work \cite{bader2023enabling}, which proposed the integration of 5G mmWave-based positioning solutions from both LoS and NLoS measurements, along with onboard motion sensors such as accelerometers, gyroscopes, and an odometer. To assess the system's robustness, our investigation will include analyzing the system under various conditions of erroneous 5G measurements. Additionally, we will conduct outage analysis by introducing artificial 5G LoS outages of varying lengths and object dynamics. This study aims to demonstrate the benefits of incorporating multipath signals in such multipath-rich environments, and how they can improve the overall robustness and reliability of the integrated positioning system. Although the used measurements were acquired from an outdoor dense urban environment, our analysis is equally important for indoor environments as both environments are filled with obstacles and multipath-rich.

Our Contribution can be summarised in the following aspects:
\begin{itemize}
  \item Conducting a comprehensive analysis of the positioning system's performance under various conditions of erroneous 5G measurements and outage scenarios, which provides insights into the system's behavior in challenging environments.
  \item For validation, a road test was conducted in downtown Toronto (Ontario, Canada) using actual OBMS measurements collected from sensors installed in the test vehicle. The road tests were carried out in a simulation environment that accurately emulates the complex urban environment of Toronto's downtown area.
  \item Examining the advantages of incorporating multipath signals for positioning in multipath-rich environments.
\end{itemize}

The remaining sections of this paper are structured as follows: In Section II, we present the system overview where we outline the key components of the positioning system under analysis. Section III describes the experimental setup used in the road test. Section IV presents the framework for robustness analysis, along with the results and discussions. Finally, in Section V, we summarize our conclusions and offer some concluding remarks.

\begin{figure*}[t]
	\centering
	\includegraphics[width=500pt]{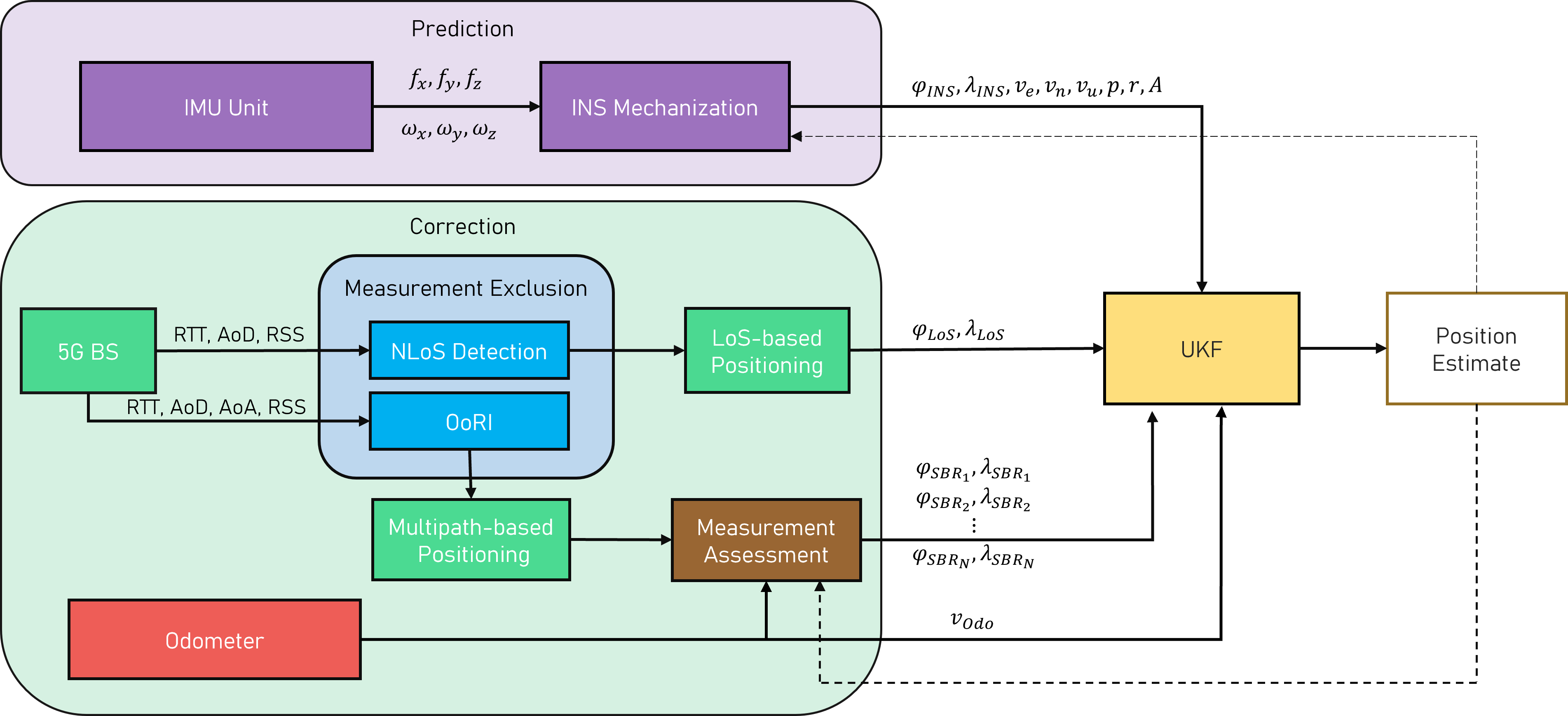}
	\DeclareGraphicsExtensions.
	\caption{Block diagram of the integrated positioning system under analysis \cite{bader2023enabling}.}
	\label{method}
\end{figure*}

\section{System Overview}
The evaluated system \cite{bader2023enabling} aims to provide an accurate and reliable positioning solution for mobile objects in restricted constrained, leveraging a fusion of 5G cellular measurements and OBMS. The system is comprised of three principal modules, namely, an inertial navigation system (INS) module, a Line-of-Sight (LoS) positioning module, and a multipath positioning module. A comprehensive depiction of the overall system architecture is presented in Fig. \ref{method}.

The INS module provides the object's position and velocity estimations obtained by integrating accelerometer and gyroscope measurements and additionally provides an estimation of the orientation of the moving object \cite{INS}. The LoS-based positioning module, on the other hand, leverages 5G LoS measurements such as Round-Trip Time (RTT) and Angle of Departure (AoD) to estimate the moving object's position and velocity \cite{LoS}. The multipath positioning module leverages reflected signals from obstructions such as buildings to estimate the position utilizing channel parameters such as time of arrival (ToA), angle of arrival (AoA), and AoD \cite{SB0}. Before the positioning estimation, a measurement exclusion process is performed to filter out Non-Line-of-Sight (NLoS) signals, allowing only LoS signals to be utilized by the LoS-based positioning module. This process is based on the distinction in distance computation between the User Equipment (UE) and the Base Station (BS) through the utilization of time-based and received signal strength-based calculations \cite{NLOS}.

Furthermore, when multipath signals are used for positioning, channel parameters are passed to an Out-of-Reflection-Region Identifier (OoRI) module \cite{OoRI}, which filters out higher-order reflections by allowing only single-bounce reflections to be passed on to the multipath positioning module. The OoRI module is based on machine learning, trained on a dataset comprising 5G channel parameters, and achieved a classification accuracy of $99.8\%$. The position computations resulting from multipath positioning undergo a second stage of validation, which is contingent upon the vehicle's motion constraints. These constraints are determined using odometer measurements and posterior estimations from the previous epoch. Finally, the position estimates from the three modules are fused using an unscented Kalman filter (UKF)\cite{ukf}, which provides a robust and accurate estimate of the mobile object's position, velocity, and orientation.

\section{Road Test Setup}
\begin{figure}[ht!]
	\centering
	\includegraphics[width=\columnwidth]{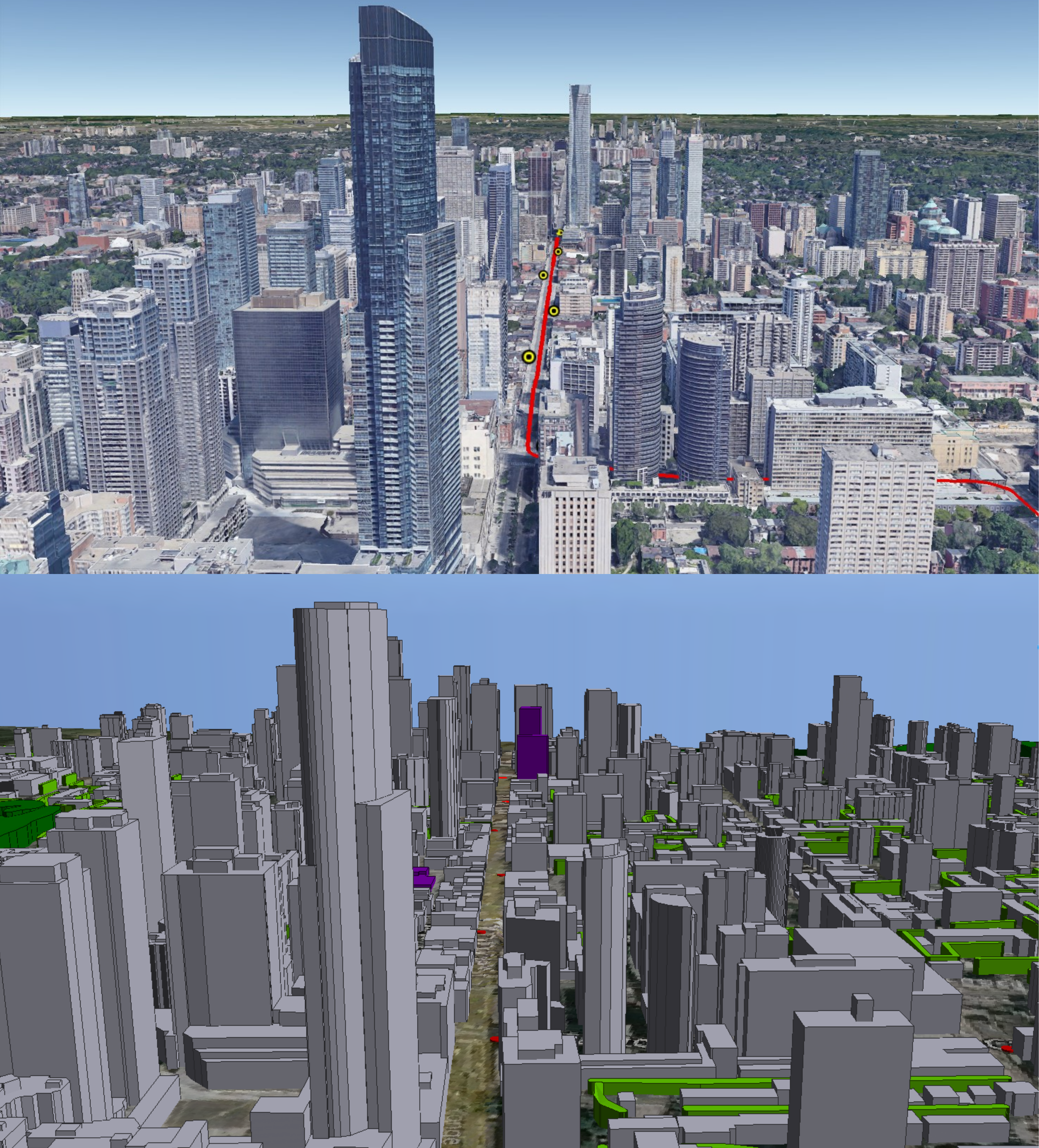}
	\DeclareGraphicsExtensions.
	\caption{Downtown Toronto, ON, Google Earth (Top) vs Siradel simulation tool (Bottom).}
	\label{GoogleEarth vs Siradel}
\end{figure}

\begin{figure}[ht!]
	\centering
	\includegraphics[width=\columnwidth]{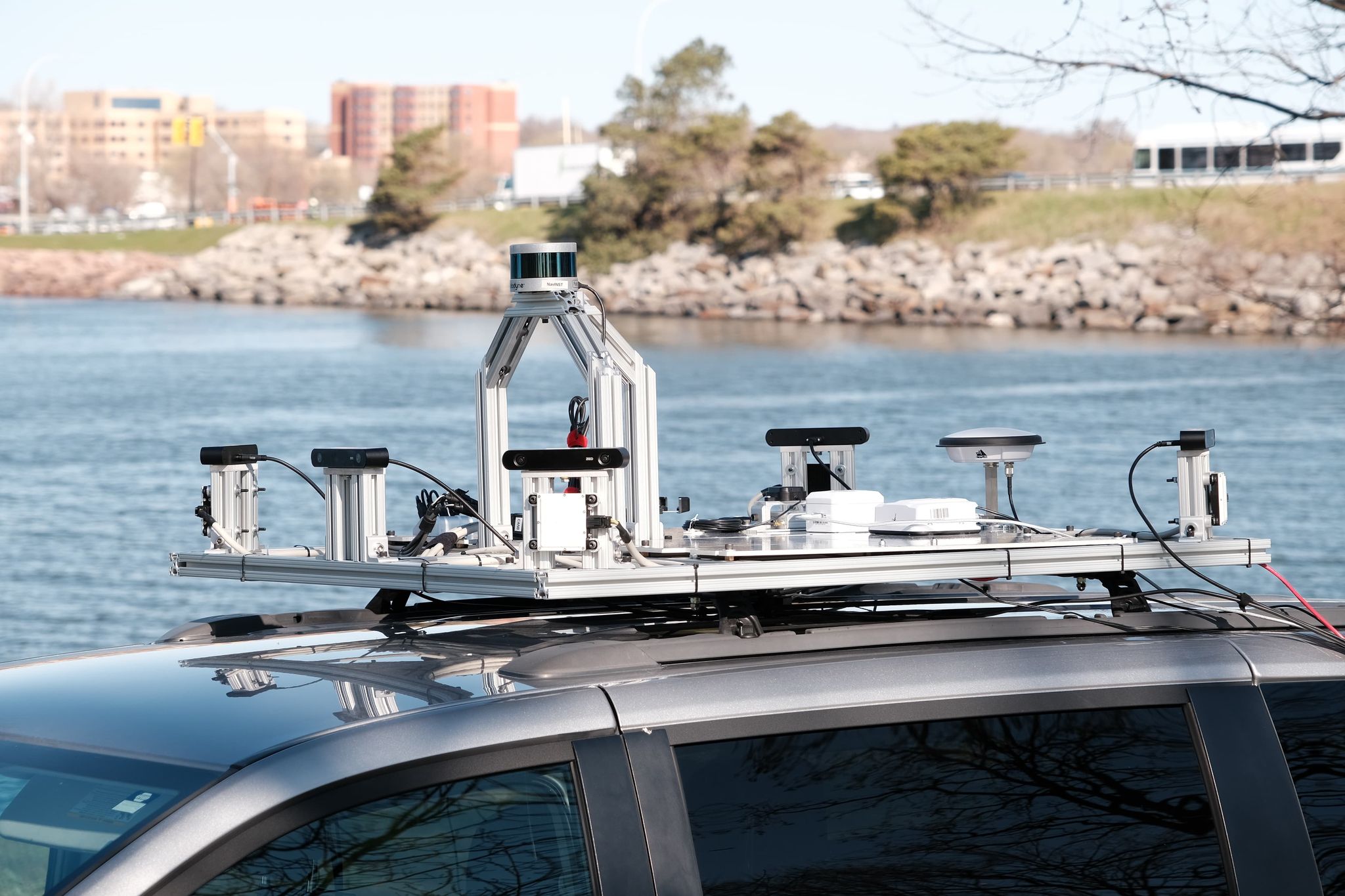}
	\DeclareGraphicsExtensions.
	\caption{NavINST testbed for multi-sensor positioning.}
	\label{testbed}
\end{figure}

For validation purposes, a quasi-real 5G simulation configuration was utilized from Siradel. As seen in Fig. \ref{GoogleEarth vs Siradel}, Siradel's 5G Channel suite comprises LiDAR-based maps of downtown city regions like Toronto to display the building structures, vegetation, and water bodies. As using a map of the actual downtown area offers accurate information on the physical features of the environment, it follows that captured multipath signals are likely to resemble realistic signal propagation. Examples of factors that affect the spread of 5G signals include building heights, roadway widths, and material compositions. The simulation tool generates positioning measurables including received signal strength (RSS), ToA, AoA, and AoD based on the UE reference positions and virtually connected BS positions by using its ray-tracing capabilities and propagation models. To collect the UE reference solution, a real car was equipped by a high-end integrated positioning system provided by Novatel as seen in Fig. \ref{testbed}. The unit features a tactical-grade IMU (KVH 1750) along with a GNSS receiver with RTK capabilities. The 3GPP's Release 16 requirements were followed, therefore BSs were spaced roughly $250$m apart throughout the driving path. The required 5G measurables were then acquired by Siradel utilizing the imported BS positions and the reference solution. The carrier frequency and bandwidth of Siradel's mmWave broadcasts were $28$ GHz and $400$ MHz, respectively. The BSs had $8\times1$ ULAs while the UE had an omnidirectional antenna.

\begin{figure}[ht!]
	\centering
	\includegraphics[width=\columnwidth]{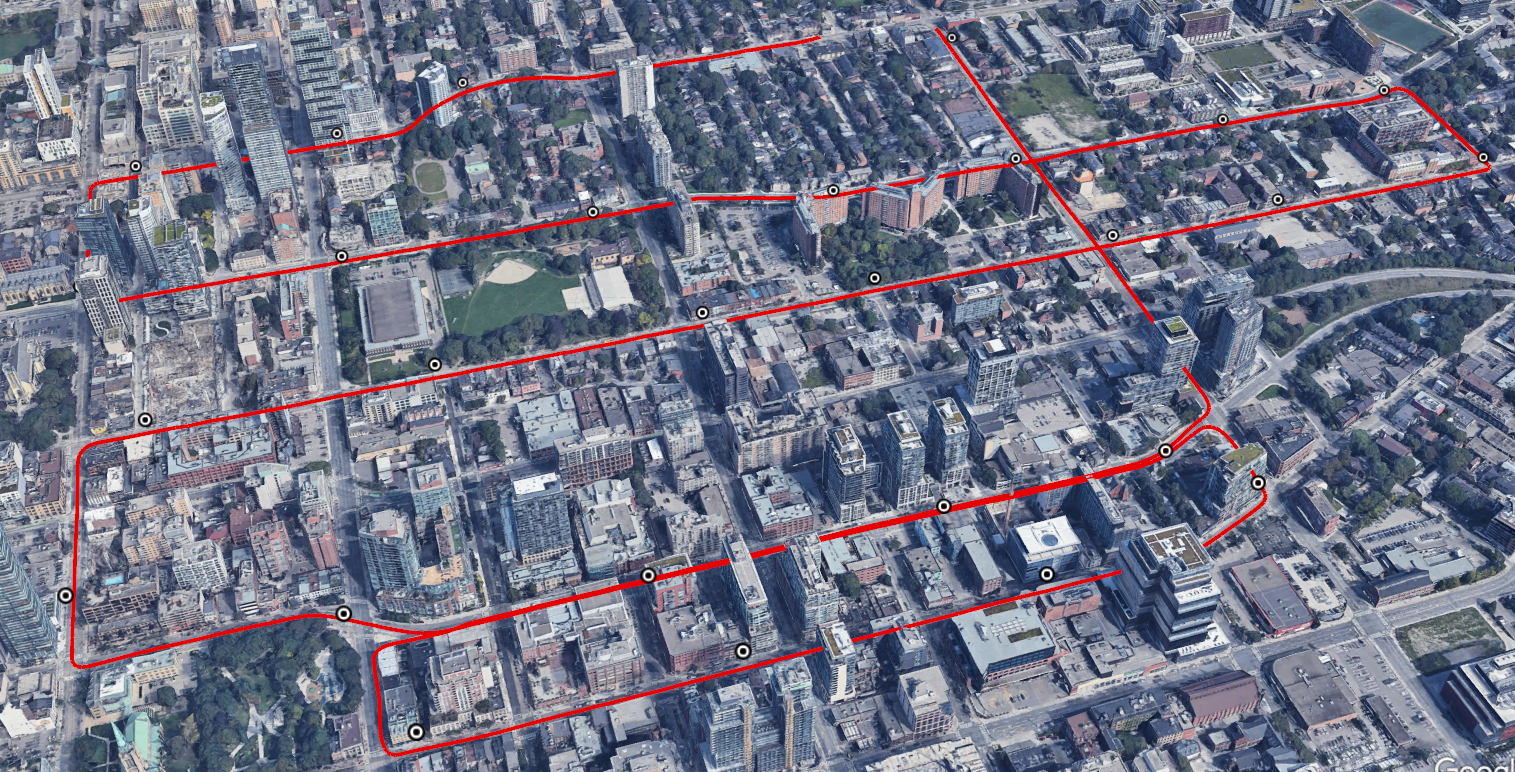}
	\DeclareGraphicsExtensions.
	\caption{Downtown Toronto Trajectory (Red), and 5G gNBs (Black circles).}
	\label{Traj}
\end{figure}

\section{Robustness Analysis Results and Discussions}
\subsection{5G-LoS Outage Analysis}
In this section, we present an analysis of the impact of introducing artificial 5G LoS outages on the 3D positioning accuracy of our proposed integrated positioning system. We evaluate the performance of the system under various outage scenarios, with varying duration, distances, speeds, and total change in azimuth (heading) angle. The objective of this analysis is to assess the system's robustness in constrained environments like indoor or urban environments, where LoS signals are subject to blockages and interruptions due to the presence of obstacles and other environmental factors. Additionally, we investigate the effect of single-bounce reflections on the system's accuracy by comparing the positioning errors with and without considering these reflections. Table \ref{outages} presents the characteristics of 10 introduced 5G LoS outages, while Table \ref{outage_results} depicts the root mean square error (RMSE) and the max error as a percentage of distance traveled during a 5G LoS outage.

\begin{table}[h]
    \caption{Artificial 5G-LoS Outages Characteristics}
    \label{outages}
    \centering
\begin{tabularx}{\columnwidth}{ccccc}
\hline \textbf{Outage \#} & $\begin{array}{c}\textbf {Duration} \\
{[\mathrm{s}]}\end{array}$ & $\begin{array}{c}\textbf {Distance} \\
{[\mathrm{m}]}\end{array}$ & $\begin{array}{c}\textbf {Avg.} \\ \textbf{Speed}\\
{[\mathrm{m} / \mathrm{s}]}\end{array}$ & $\begin{array}{c}\Delta\\ \textbf {Azimuth} \\
{[\mathrm{deg}]}\end{array}$ \\[3pt]
\hline \\
1 & 20 & 196 & 9.8 & 72 \\[3pt]
2 & 20 & 174 & 8.7 & 3.1 \\[3pt]
3 & 20 & 3.4 & 0.17 & 1.3 \\[3pt]
4 & 40 & 374 & 9.4 & 152 \\[3pt]
5 & 40 & 192 & 4.7 & 14.7 \\[3pt]
6 & 40 & 25.6 & 0.6 & 3.8 \\[3pt]
7 & 60 & 547 & 9.1 & 156 \\[3pt]
8 & 200 & 1137 & 5.6 & 300 \\[3pt]
9 & 400 & 2615 & 6.5 & 674 \\[3pt]
10 & 1000 & 6255 & 6.3 & 29324 \\[3pt]
\hline
\end{tabularx}
\end{table}

\begin{table}[h]
    \caption{3D Positioning Errors With and Without SBRs}
    \label{outage_results}
    \centering
\begin{tabularx}{\columnwidth}{ccccc}
\hline\\ & \multicolumn{2}{c}{\hspace{20pt}\textbf {Without SBRs}\hspace{20pt}} & \multicolumn{2}{c}{\hspace{20pt}\textbf {With SBRs}\hspace{20pt}} \\[0.3pt]
\cline { 2 - 5 }\\[0.3pt] \hspace{20pt}\textbf {Outage \#}\hspace{20pt} & \textbf {  RMS } & \% & \textbf { RMS } & \% \\[3pt]
\hline \\
1 & 0.82 & 0.73 & 0.07 & 0.13 \\ [3pt]
2 & 0.90 & 1.10 & 0.04 & 0.11 \\[3pt]
3 & 0.94 & 40.4 & 0.01 & 0.40 \\[3pt]
4 & 2.11 & 1.23 & 0.10 & 0.14 \\[3pt]
5 & 2.15 & 1.86 & 0.03 & 0.11 \\[3pt]
6 & 0.78 & 6.26 & 0.01 & 0.06 \\[3pt]
7 & 3.74 & 1.20 & 0.08 & 0.09 \\[3pt]
8 & 6.32 & 0.79 & 0.10 & 0.07 \\[3pt]
9 & 9.36 & 0.63 & 0.26 & 0.09 \\[3pt]
10 & 26.1 & 1.37 & 12.5 & 0.77 \\[3pt]
\hline
\end{tabularx}
\end{table}

The outages' duration ranges from $20$ seconds to $1000$ seconds, and their distances range from 3.4 meters to 6255 meters. The average speed of the object during the outages ranges from $0.17$ m/s to $9.8$ m/s, while the total cumulative change in azimuth angle ranges from $3.3$ degrees to $29324$ degrees. Overall, it can be seen that the presence of SBRs significantly reduces the positioning error in all cases. The observation that the positioning error tends to increase with the duration of the outage, with the exception of outage 6, warrants further analysis. Several factors may have contributed to this outcome. Notably, errors in the estimation of the azimuth angle could lead to inaccuracies in the determination of the object's position, which, in turn, is modulated by the object's speed which is referred to as non-stationary error \cite{ProfBook}.

\begin{figure}[ht!]
	\centering
	\includegraphics[width=\columnwidth]{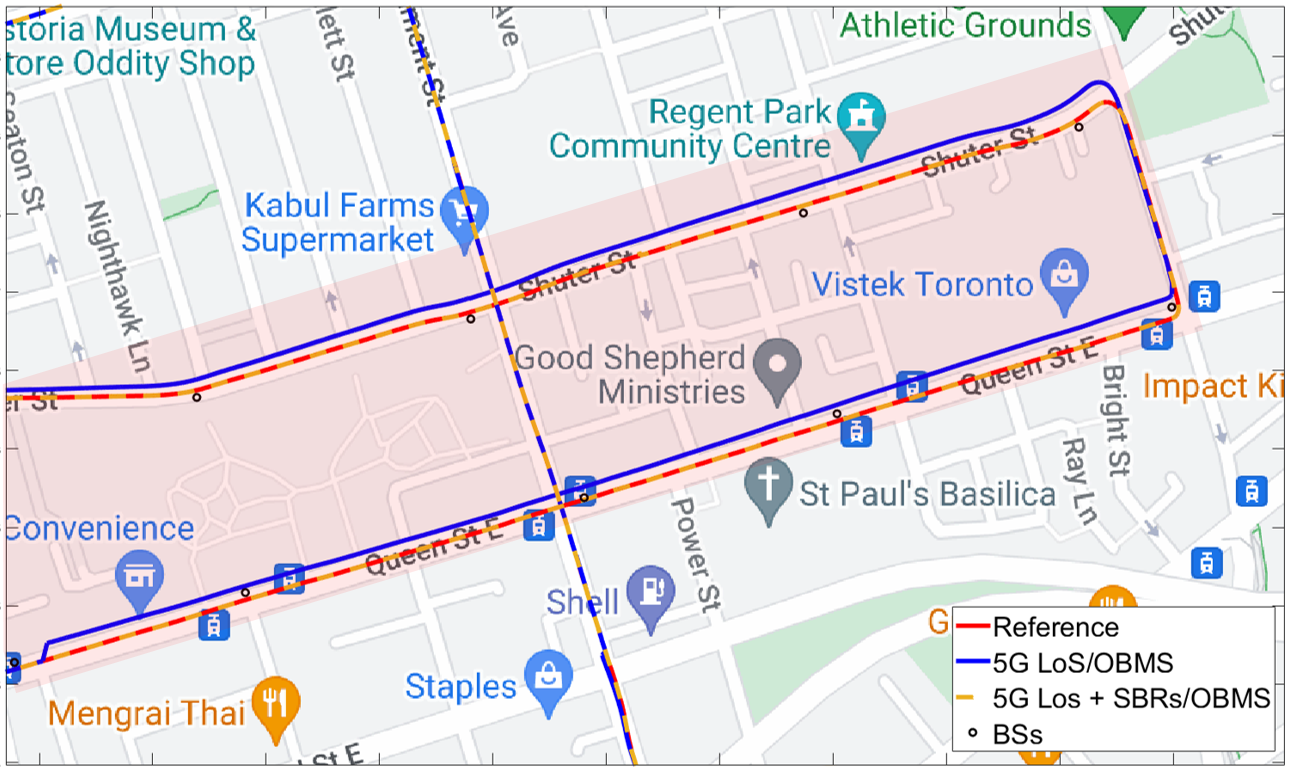}
	\DeclareGraphicsExtensions.
	\caption{Close-up scenario of 5G positioning solution during LoS outage (Shaded area) with and without SBRs.}
	\label{outage9}
\end{figure}

Fig. \ref{outage9} provides a close-up view of the positioning solution during outage \#9, both before and after incorporating SBRs. The plot demonstrates that, as a result of the drifting errors of INS, the positioning solution without multipath assistance drifts away from the reference solution. On the other hand, the multipath-assisted measurements continue to closely follow the reference solution.

\subsection{Effect of Noisy Range and Angle Measurements}
\begin{figure}[t!]
	\centering
	\includegraphics[trim=110pt 20pt 120pt 30pt,clip,width=\columnwidth]{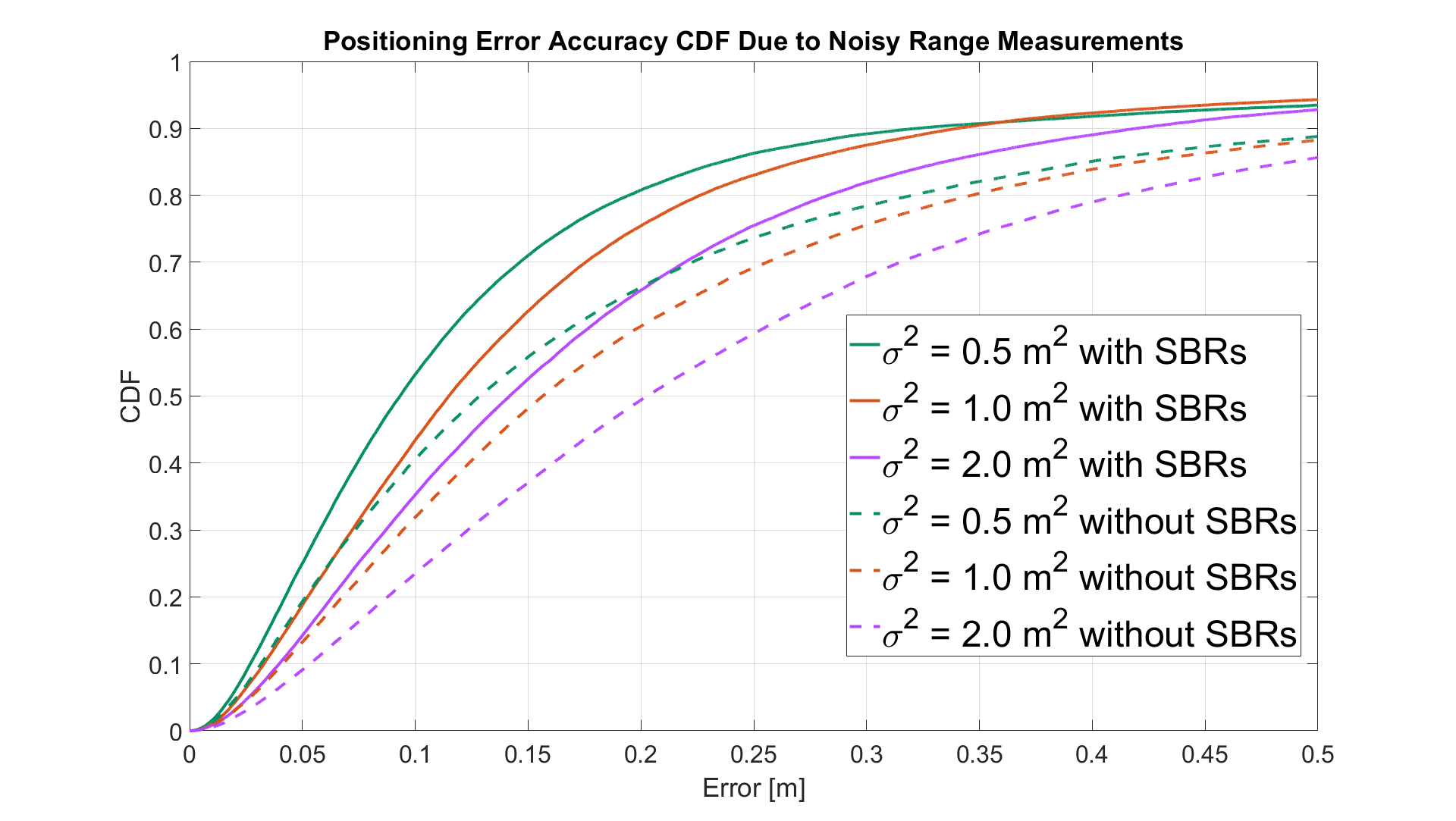}
	\DeclareGraphicsExtensions.
	\caption{Positioning CDF due to noisy range measurements.}
	\label{range}
\end{figure}

In this section, we introduce normally distributed errors in the angle and range measurements with varying variances to analyze the robustness of the positioning system before and after incorporating multipath signals. Figure. \ref{range} showcases the cumulative distribution function (CDF) of the positioning accuracy under noisy range measurements of variances $0.5~\text{m}^2$, $1~\text{m}^2$, and $2~\text{m}^2$. In general, the utilization of multipath measurements appears to result in superior positioning accuracy. Specifically, it has been observed that the positioning accuracy with a variance of $1~\text{m}^2$ is consistently higher than that achieved with a variance of $0.5$ $\text{m}^2$ when relying solely on LoS signals. 

\begin{figure}[t!]
	\centering
	\includegraphics[trim=185pt 90pt 130pt 60pt,clip,width=\columnwidth]{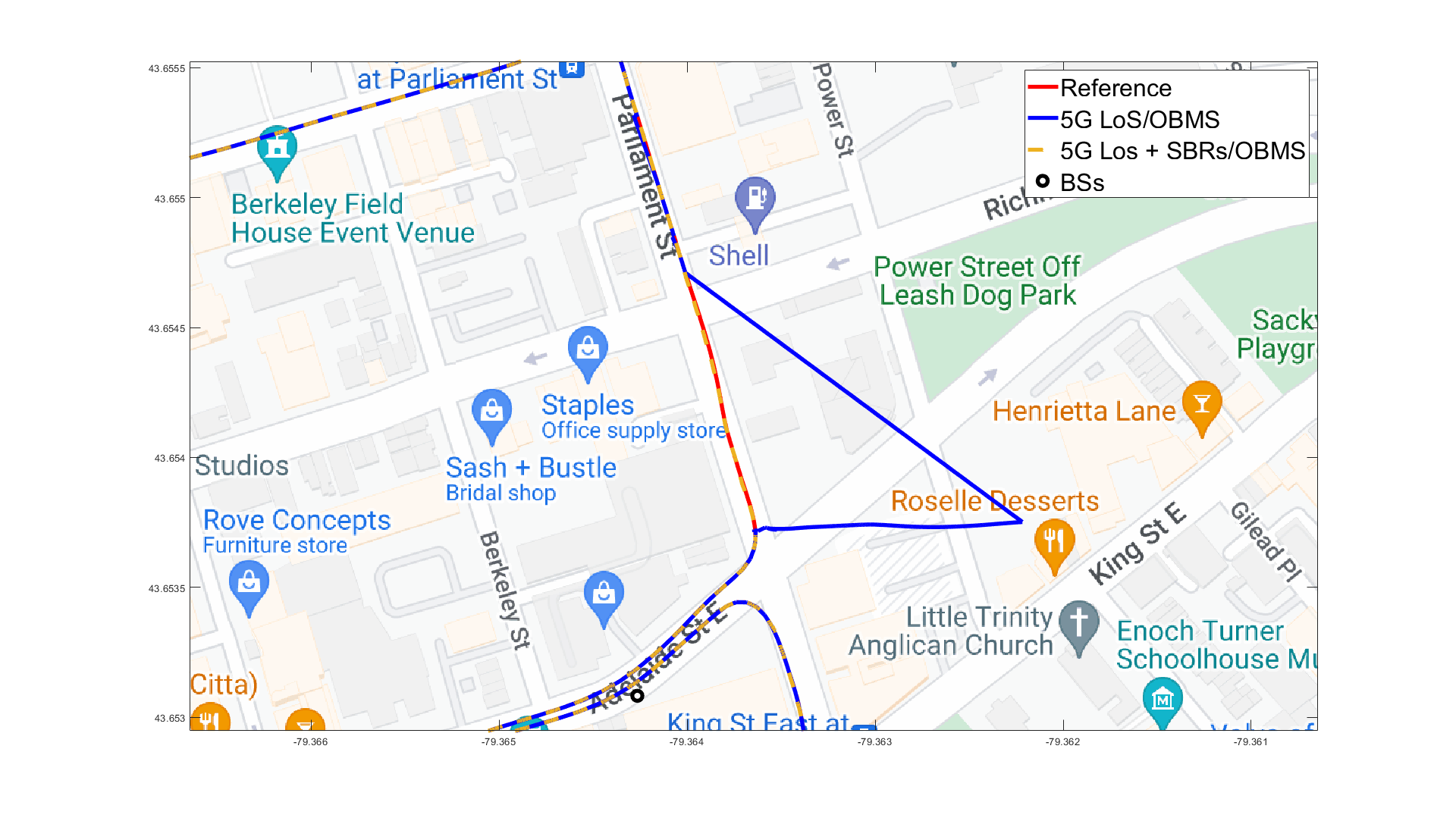}
	\DeclareGraphicsExtensions.
	\caption{Close-up scenario of the positioning solution with and without SBRs under noisy range measurements.}
	\label{range_map1}
\end{figure}

\begin{figure}[t!]
	\centering
	\includegraphics[trim=185pt 90pt 130pt 60pt,clip,width=\columnwidth]{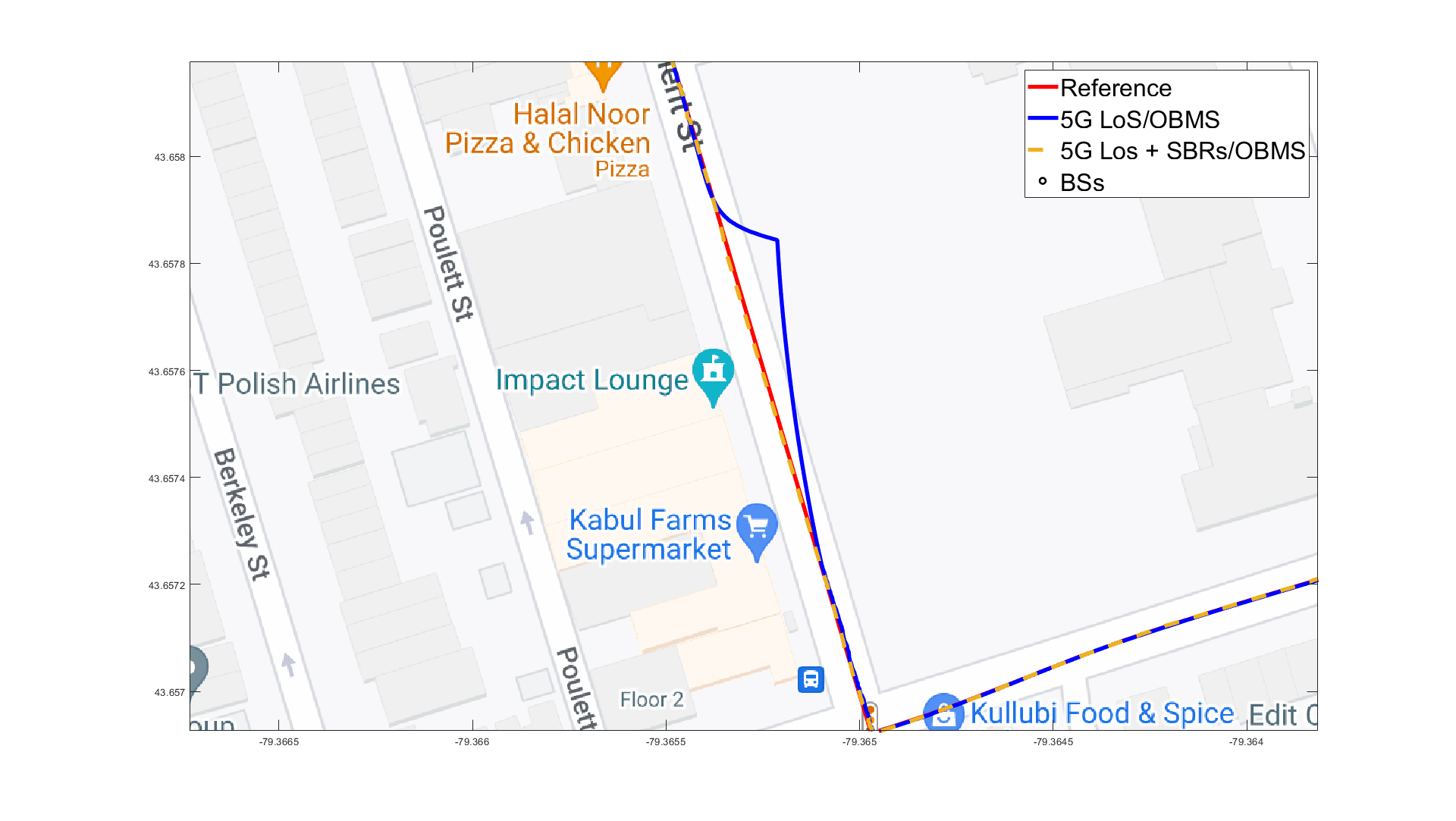}
	\DeclareGraphicsExtensions.
	\caption{Close-up scenario of the positioning solution with and without SBRs under noisy range measurements.}
	\label{range_map2}
\end{figure}

Figs. \ref{range_map1} and \ref{range_map2} depict the positioning solution prior to and after the integration of SBRs under the influence of noisy range measurements with a variance of $2~\text{m}^2$. It is observed that the presence of noise in the measurements during the 5G LoS outage, represented by the obstruction of the BS, leads to large errors in the positioning solution. This can be potentially attributed to erroneous corrections to the INS mechanization process, compounded by a complete loss of the LoS measurements. Consequently, the computed position based on dead reckoning would be subject to substantial bias until an LoS signal becomes available. The integration with SBRsled to a notable enhancement in the positioning solution. This improvement can be attributed to the augmented redundancy of measurements thanks to the increased availability of SBRs compared to LoS communication. As a result, the solution was able to track the reference solution more accurately. This highlights the importance of incorporating redundant measurements, particularly in scenarios where the primary measurement source, such as the LoS signal, may not be available.

\begin{figure}[t!]
	\centering
	\includegraphics[trim=110pt 20pt 120pt 30pt,clip,width=\columnwidth]{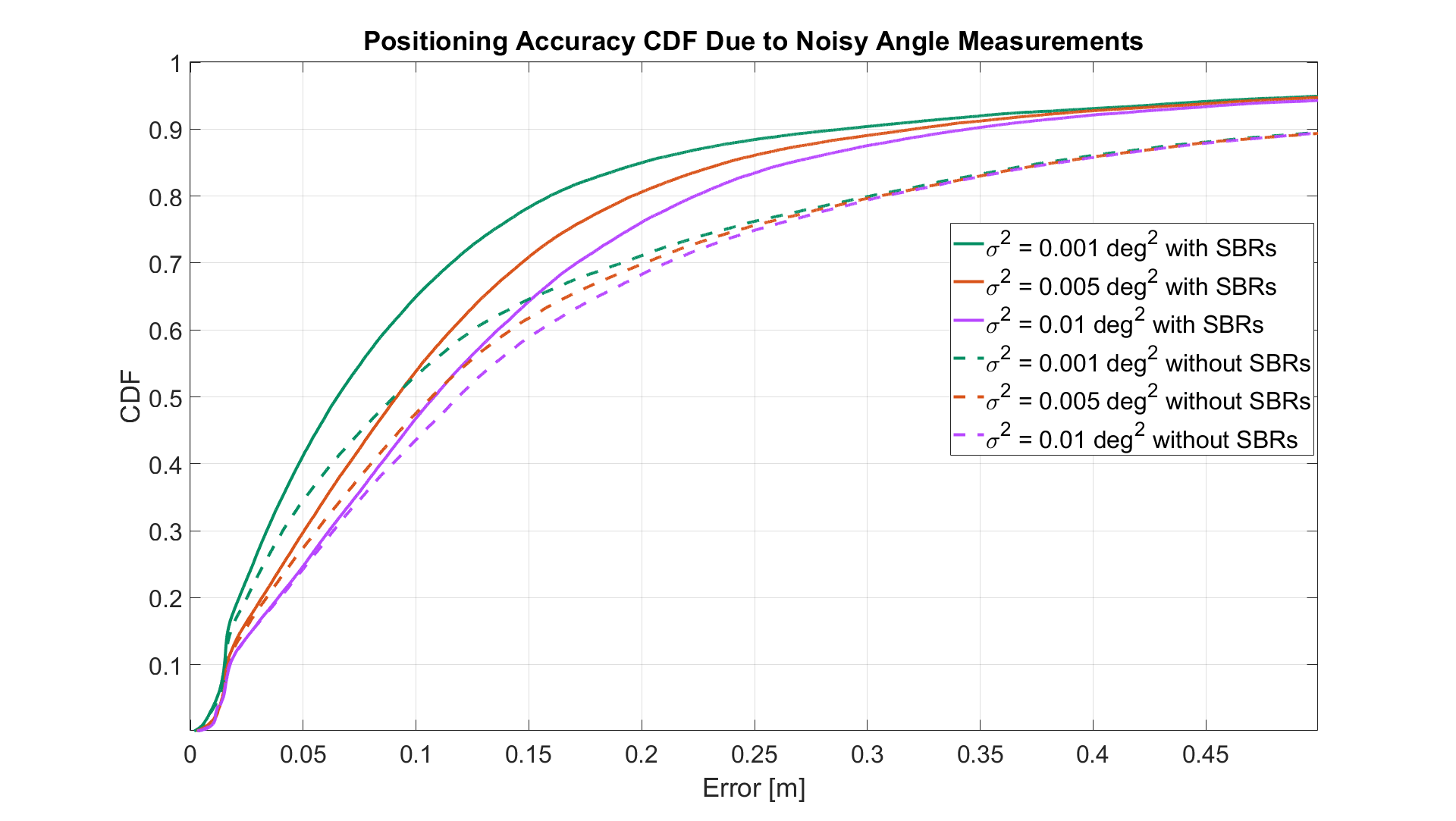}
	\DeclareGraphicsExtensions.
	\caption{Positioning CDF due to noisy angle measurements.}
	\label{angle}
\end{figure}

\begin{figure}[t!]
	\centering
	\includegraphics[trim=185pt 90pt 130pt 60pt,clip,width=\columnwidth]{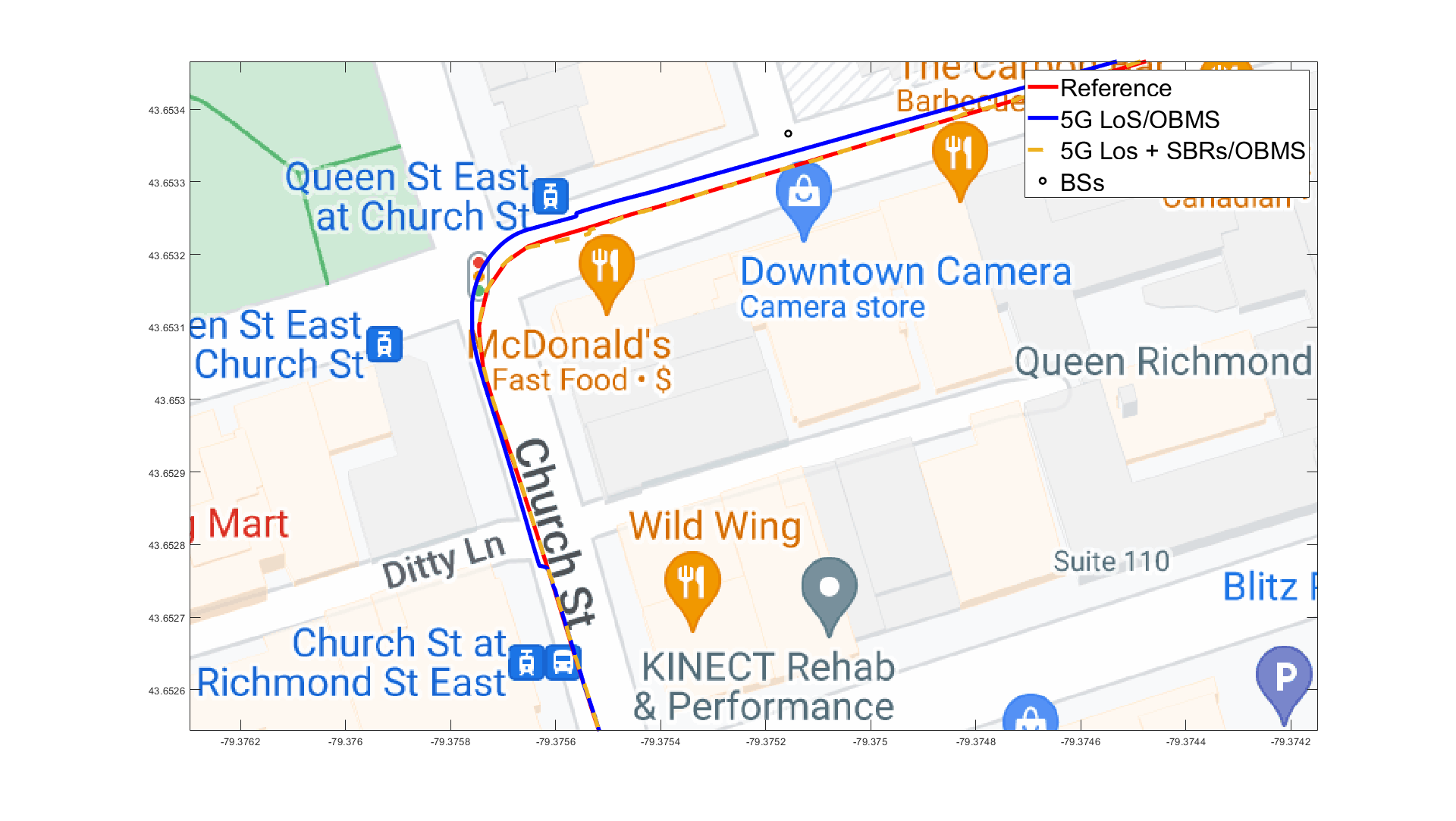}
	\DeclareGraphicsExtensions.
	\caption{Close-up scenario of the positioning solution with and without SBRs under noisy angle measurements.}
	\label{angle_map1}
\end{figure}

Fig. \ref{angle}, in contrast, illustrates the CDF of positioning accuracy affected by noisy angle measurements. It is worth noting that the computation of position using multipath signals was impacted by two types of noisy measurements, namely, AoA and AoD, while only AoA measurements affected the LoS signals. Despite this, the use of SBRs resulted in superior positioning accuracy. Examining the sub-$20$ cm level of accuracy, it can be observed that utilizing multipath signals with a variance of $0.01~\text{deg}^2$ produced better positioning accuracy than utilizing LoS signals with a variance of $0.001~\text{deg}^2$. Fig. \ref{angle_map1} demonstrates the effect of noisy angle measurements with a variance of $0.05~\text{deg}^2$ on the positioning solution. As previously observed in Figs. \ref{range_map1} and \ref{range_map2}, the integration of SBRs successfully reduced the large errors associated with noisy measurements during LoS outages.

\section{Conclusion}
In conclusion, this study has demonstrated the potential benefits of utilizing multipath signals in 5G-based indoor positioning systems. The high accuracy of 5G mmWave range and angle measurements, combined with its ability to resolve multipath signals, presents a promising opportunity for achieving accurate and reliable indoor positioning. Through our experiments, we have shown that incorporating multipath-based measurements into our positioning system can significantly reduce RMS errors during 5G signal outages, improving the robustness and accuracy of the system. Moreover, the system was also evaluated under noisy range and angle measurements with varying variances. The results demonstrate that incorporating multipath measurements leads to improved positioning accuracy, even when the system is affected by two noisy measurements instead of one, compared to the system that does not use multipath assistance. These findings suggest that incorporating multipath signals into 5G-based indoor positioning systems can enhance their performance and open up new opportunities for a wide range of applications. Further research is required to explore the full potential of multipath-based positioning systems in indoor environments utilizing higher-order reflections.

\bibliographystyle{IEEEtran}
\bibliography{References}
\end{document}

%% file: Preamble.tex
\IEEEoverridecommandlockouts

\usepackage{cite}
\usepackage{amsmath,amssymb,amsfonts}

\usepackage{algorithmic}
\usepackage[ruled, linesnumbered, vlined]{algorithm2e} 

\usepackage{graphicx}
\usepackage{textcomp}
\usepackage{xcolor}
\graphicspath{ {./images/} }
\usepackage{booktabs}
\usepackage{lscape}
\usepackage{adjustbox}

\usepackage{caption}
\usepackage{subcaption}
\usepackage[left=1.62cm,right=1.62cm,top=0.8in]{geometry}
\usepackage{lipsum} 
\usepackage{flafter} 
\usepackage{color,soul} 
\usepackage{tabularx,booktabs} 
\newcolumntype{C}{>{\centering\arraybackslash}X} 
\usepackage[bookmarks=false,hidelinks,draft]{hyperref} 
\usepackage{scalerel} 
\usepackage{tikz} 

\renewcommand{\baselinestretch}{1.01}
\def\BibTeX{{\rm B\kern-.05em{\sc i\kern-.025em b}\kern-.08em T\kern-.1667em\lower.7ex\hbox{E}\kern-.125emX}}
\usepackage{stfloats}

\usetikzlibrary{svg.path}
\definecolor{orcidlogocol}{HTML}{A6CE39}
\tikzset{
	orcidlogo/.pic={
		\fill[orcidlogocol] svg{M256,128c0,70.7-57.3,128-128,128C57.3,256,0,198.7,0,128C0,57.3,57.3,0,128,0C198.7,0,256,57.3,256,128z};
		\fill[white] svg{M86.3,186.2H70.9V79.1h15.4v48.4V186.2z}
		svg{M108.9,79.1h41.6c39.6,0,57,28.3,57,53.6c0,27.5-21.5,53.6-56.8,53.6h-41.8V79.1z M124.3,172.4h24.5c34.9,0,42.9-26.5,42.9-39.7c0-21.5-13.7-39.7-43.7-39.7h-23.7V172.4z}
		svg{M88.7,56.8c0,5.5-4.5,10.1-10.1,10.1c-5.6,0-10.1-4.6-10.1-10.1c0-5.6,4.5-10.1,10.1-10.1C84.2,46.7,88.7,51.3,88.7,56.8z};
	}
}
\newcommand{\orcidicon}[1]{\href{https://orcid.org/#1}{\mbox{\scalerel*{
				\begin{tikzpicture}[yscale=-1,transform shape]
				\pic{orcidlogo};
				\end{tikzpicture}
			}{|}}}}

\newcommand{\ignore}[1]{}

\newcommand{\linebreakand}{%
\end{@IEEEauthorhalign}
\hfill\mbox{}\par
\mbox{}\hfill\hspace*{-1cm}\begin{@IEEEauthorhalign} 
}

